# Fluorescence microscopy beyond the ballistic regime by ultrasound pulse guided digital phase conjugation

Meng Cui[*], Ke Si and Reto Fiolka

*Howard Hughes Medical Institute, Janelia Farm Research Campus,
19700 Helix Drive, Ashburn, Virginia, 20147, USA*

[*]Correspondence and requests for materials should be addressed to Meng Cui, 571-209-4136
cuim@janelia.hhmi.org

**Abstract**

Fluorescence microscopy has revolutionized biomedical research over the past three decades. Its high molecular specificity and unrivaled single molecule level sensitivity have enabled breakthroughs in a variety of research fields. For *in vivo* applications, its major limitation is the superficial imaging depth as random scattering in biological tissues causes exponential attenuation of the ballistic component of a light wave. Here we present fluorescence microscopy beyond the ballistic regime by combining single cycle pulsed ultrasound modulation and digital optical phase conjugation. We demonstrate near isotropic 3D localized sound-light interaction with an imaging depth as high as thirteen scattering path lengths. With the exceptionally high optical gain provided by the digital optical phase conjugation system, we can deliver sufficient optical power to a focus inside highly scattering media for not only fluorescence microscopy but also a variety of linear and nonlinear spectroscopy measurements. This technology paves the way for many important applications in both fundamental biology research and clinical studies.

The capability of *in vivo* fluorescence microscopy has expanded rapidly over the past few years[1-4]. Despite the progress in spatial resolution[1] and imaging speed[2, 4], the achievable imaging depth in live samples remains very limited[5, 6], which has hindered the progress of many research fields[3, 7, 8]. The bottle neck is that only the ballistic component of a light wave has been utilized for imaging, which experiences exponential attenuation due to random scattering in tissues[5, 6]. Here we report fluorescence microscopy beyond the ballistic regime, which combines single cycle pulsed ultrasound modulation and digital optical phase conjugation[9]. Utilizing single cycle focused ultrasound, we can generate a near isotropic 3D confined sound-light modulation volume. Through digital optical phase conjugation, we can focus sufficient light energy at will to an arbitrary point inside random scattering media, determined by the timing and the propagation direction of the focused sound pulse. This technology paves the way for not only deep tissue fluorescence microscopy but also a variety of linear and nonlinear spectroscopy measurements at unprecedented depths for both fundamental research and clinical applications.

Controlling the propagation of an optical wave has been an interesting and important subject in many research fields[10-16]. In optical microscopy, objective lenses have been employed to convert a collimated beam into an optical focus or vice versa. However such a simple conversion breaks down in highly scattering media, as the aberration and random scattering gradually disrupt the originally converging or diverging wavefront. Consequently, optical microscopy has been limited to a few scattering path lengths[6]. Deeper into tissues, the wavefront is randomized and consequently no focus can be formed. To reveal the optical information hidden inside highly scattering tissues, more sophisticated wavefront control is required.

In principle, if one can reverse both the propagation direction and wavefront of the optical wave originating from a point (i.e. a guide star) inside turbid media, one can form an optical focus at the original point regardless the thickness of the turbid media, a process known as optical phase conjugation (OPC)[9, 11, 12, 17]. For imaging, the challenging task is how to freely place a guide star at arbitrary locations inside turbid media. Recently, it has been proposed to use a sound wave to modulate light as a means to create a guide star[18]. As the scattering of sound waves in tissues is negligible in comparison with light[19], the guide star can be placed at depth far beyond the ballistic regime of light. However, for practical fluorescence microscopy, there are two remaining challenges. First, sound and light are both propagating wave in tissues. Even with focused ultrasound, their interaction volume is not 3D confined. Second, given a 3D confined interaction volume, the amount of light that is sound modulated within highly scattering medium is very small, as only a small fraction of diffused light can propagate through the confined interaction volume and the light diffraction efficiency by sound in tissues is rather low (~ 1%). Thus for practical imaging applications and potential spectroscopy measurements in deep tissues, we need tremendous optical gain ($> 10^3$) for the phase conjugation beam, which cannot be readily provided by a conventional phase conjugation system using photorefractive crystals[10, 20].

Here, we report fluorescence microscopy beyond the ballistic regime with < 40 microns spatial resolution at an imaging depth of 13 scattering path lengths. Different from the previous report, we use single cycle focused ultrasound pulses and tightly synchronized laser pulses to achieve a near isotropic 3D confined interaction volume. The pulsed light and pulsed sound wave are precisely synchronized such that the light wave illuminates the sample only when the single cycle ultrasound pulse propagates to its spatial focus. In such a way, the sound modulation zone is confined in the transverse direction to < 40 microns by the sound focusing element and in the axial direction to < 40 microns by the temporal profile of the single cycle sound pulse convolved with the temporal profile of the laser pulse. With such a scheme, we achieved a near isotropic 3D confined sound light interaction volume. Photorefractive crystals typically used in conventional OPC experiments can only provide moderate optical gain[21] and the recorded hologram is erased during readout[18, 20]. To provide sufficient and durable optical power for fluorescence microscopy, we employed digital optical phase conjugation[9] (DOPC) to perform phase conjugation. The DOPC system operates in two steps. In the first step, it performs digital holography with a CMOS sensor to measure the wavefront of the sound modulated light. In the second step, the measured phase profile is sign reversed and displayed on a high resolution spatial light modulator (SLM) that is precisely aligned with respect to the CMOS sensor such that their pixels form mirror images through an optical beam splitter. A laser beam illuminates the SLM and becomes the phase conjugation beam. As such, the power of the phase conjugation beam can be freely adjusted by changing the power of the incident laser beam on the SLM and exceptionally high optical gain can be achieved.

Figure 1 (*a*) illustrates schematically the operation of the fluorescence imaging system. A high frequency focused ultrasound transducer launches a single cycle pulse into the sample. A short laser pulse illuminates the sample only when the sound pulse travels to its focus. Only the light that propagates through the 3D confined modulation zone is frequency shifted. The wavefront of the frequency shifted light is recorded by the DOPC system using heterodyne phase stepping interferometry. To measure the fluorescence signal, the DOPC system sends out the phase conjugation beam that precisely propagates to the sound focus. A fluorescence detector measures the power of the emitted fluorescence light. To form a fluorescence image, the entire process is repeated as the acoustic focus is raster scanned inside the sample by translating the ultrasound transducer with a three-axis motorized stage.

Figure 1 (*b*) shows the setup of the fluorescence imaging system. The laser source is a Q-switched green laser pumped Ti:sapphire laser (Photonics Industries, NY) with 778 nm center wavelength, 10 kHz repetition rate and 20 ns pulse duration. The laser power is controlled by a half wave plate and a polarizing beam splitter. The laser output was split into two beams. One beam travels through a beam expander and enters the DOPC system. During wavefront recording, it serves as the reference beam for interferometry. During the fluorescence excitation, this beam illuminates the SLM and becomes the phase conjugation beam. The other beam is modulated by an acousto-optic modulator (AOM). The frequency shifted component travels through a beam expander and illuminates the sample during wavefront recording. The sample is housed inside a water chamber with optical windows. An ultrasound transducer is mounted on a three-axis translation stage and the sound wave enters the sample from below. The phase conjugation beam is filtered by a band-pass filter before entering the water chamber for fluorescence excitation. The fluorescence emission is filtered by a long-pass filter before entering the fluorescence power detector. The timing and synchronization layout is shown in Supplementary Figure 1 and the details are discussed in Methods.

To measure the point spread function (PSF) of the system, we dispersed 6 microns diameter fluorescence beads in agar and sandwiched the fluorescence sample between two 2 mm thick scattering tissue phantoms with a scattering coefficient ($\mu_s$) 6.4±0.2/mm. The scattering coefficient measurement is discussed in Methods. The tissue phantom was parallel to the yz plane in Fig. 1 (*b*). Since there is no tissue phantom between the detector (Andor iXon 3 888 CCD) and the fluorescence beads, we could directly image individual fluorescence beads onto the detector. We raster scanned the ultrasound focus around the location of a single bead and performed DOPC at each position to excite fluorescence. The fluorescence emission from the single bead was recorded at each scan position. Figure 2 (*a*) shows the measured transverse (yz plane) PSF. The sampling step size is 15 microns. We used bicubic interpolation to resample the image, as shown in Fig. 2 (*c*). We used a Gaussian curve to fit the transverse PSF in Fig. 2 (*a*) and the fitting shows a FWHM of 38.6 microns in y direction (Fig. 2 (*e*)) and 37.9 microns in z direction (Fig. 2 (*f*)). Figure 2 (*b*) shows the measured axial (xy plane) PSF. The sampling step size is 15 microns in y direction and 80 microns in x direction. The resampled image is shown in Fig. 2 (*d*). Figure 2 (*g*) shows the Gaussian fitting along the x axis with a FWHM of 263 microns. Supplementary Figure 2 (*a*) shows one of the phase patterns ($p(y,z)$) used during the PSF measurement for OPC beam generation. In supplementary Fig. 2 (*b*) and (*c*), the autocorrelation of $\exp(ip(y,z))$ along the y and z axis is shown, respectively. Despite the apparent randomness, the phase profile is deterministic and highly sensitive on the initial condition (i.e. the position of the acoustic focus within the sample). Experimentally we found that the displacement of the phase pattern by a few pixels on the SLM made the DOPC ineffective.

Although without wavefront control the input laser light becomes randomized by the scattering medium, it can still excite fluorescence, which is the background signal in our measurements. Supplementary Figure 2 (*d*) shows the fluorescence signal from one bead illuminated by an OPC beam and (*e*) shows the fluorescence signal after the phase pattern is translated by 30 pixels both in y and z directions on the SLM (The translation makes OPC ineffective while the total illumination power at the sample remains constant). To remove the background signal, experimentally we measured the fluorescence signals with and without translating the phase pattern and the difference between the two signals was used to represent the fluorescence signal at the sound modulation position.

To demonstrate the fluorescence imaging capability, we used a glass micropipette to manually create an array of 60 microns diameter holes with 120 microns spacing in agar and injected fluorescence beads inside the holes to create a fluorescence pattern. A direct imaging is shown in Fig. 3 (*a*). The fluorescence hole array was then surrounded with 2 mm thick tissue phantoms. Figure 3 (*b*) shows the fluorescence image of the hole array with tissue phantoms around it. Due to random scattering, the image diffused to ~ 2 mm in diameter and the structure information is completely lost. We raster scanned the position of the acoustic focus and performed DOPC based fluorescence excitation. The fluorescence power was recorded by summing up all the pixels of the detector at each scanning position and the raw data is shown in Fig. 3 (*c*). The scanning step size was set to 30 microns, sufficient for resolving the fluorescence hole array. The raw data was again resampled with bicubic interpolation, as shown in Fig. 3 (*d*).

In our experiments, we used focused single cycle 50 MHz ultrasound pulse to achieve < 40 microns near isotropic 3D confined modulation zone. The dependence of the modulation zone on experimental parameters is analyzed in Supplementary discussion. For applications requiring higher spatial resolution, higher frequency ultrasound transducer can be employed to shrink the modulation zone, as both the transverse and axial dimensions of the modulation zone are proportional to the sound wavelength. In the fluorescence imaging experiments, we used one-photon fluorescence excitation, for which the fluorescence excitation is not 3D confined. The background and the out-of-focus excitations reduce the achievable signal to noise ratio (SNR). Much stronger fluorescence confinement and thus reduced background could be achieved using two-photon excitation. A picosecond regenerative amplifier might be used as the laser source. In addition to suppressing the background and the out-of-focus fluorescence excitations, two-photon excitation can further reduce the PSF by ~ $\sqrt{2}$ in each dimension due to the square dependence of the fluorescence excitation to light intensity.

Previous study[14] suggests that the achievable focus to background ratio (FBR) is proportional to $N_{pixel}/N_{mode}$, where $N_{pixel}$ is the number of independently controlled phase pixels on the SLM and $N_{mode}$ is the number of optical modes at the sound modulation zone. There are 1920 x 1080 pixels on the SLM. In the current implementations, we arranged the optical signal collection such that one speckle occupies ~2x2 pixels on the SLM (Supplementary Fig. 2 (*a*)-(*c*)) to avoid the phase errors due to pixel-to-pixel phase coupling. In addition, the SLM has limited filling factor and diffraction efficiency, and temporal phase fluctuation[22]. Employing an SLM with less pixel-to-pixel coupling, higher filling factor and diffraction efficiency, and lower temporal phase fluctuation can potentially improve FBR by more than one order of magnitude. Moreover, the sound modulation zone can be shrunken by using higher frequency sound transducers, reducing $N_{mode}$ and further improving FBR.

In conclusion, we report fluorescence microscopy beyond the ballistic regime. Using focused single cycle ultrasound pulse and synchronized laser pulse illumination, we achieved a 3D confined sound modulation zone. With the exceptionally high optical gain provided by the DOPC system, we have sufficient optical power for fluorescence microscopy. In the demonstration, we achieved < 40 microns resolution at an imaging depth of 13 scattering path lengths. The spatial resolution can be further improved with higher frequency ultrasound transducers. Combining the current system with picosecond regenerative amplifier for two-photon excitation can further improve the spatial resolution while suppressing background and out-of-focus fluorescence excitation. With the capability of focusing sufficient optical power inside random scattering media, our system can be used for not only fluorescence microscopy but also a variety of linear and nonlinear spectroscopy measurements, which is expected to find numerous important biomedical applications.

## Methods

**Timing and Synchronization**

Supplementary Figure 1 shows the diagram of the timing and synchronization. A delay generator (DG1, Stanford Research DG645) was used as the master clock of the system. It output a 10 MHz TTL pulse train to synchronize an arbitrary waveform generator (AWG, Tektronix AFG3252) and the other delay generator (DG2, Stanford research DG535). DG1 sent out three 10 kHz pulse trains to trigger the Q-switched laser, the arbitrary wavefront generator, and DG2 that controlled the exposure of the CMOS camera. AWG output a 20 ns duration single cycle sinusoidal signal (Supplementary Fig. 3 (*c*)) to drive the ultrasound transducer. The ultrasound frequency was centered at 50,005,000 Hz and the pulse repetition rate was 10 kHz. Essentially, the signal was a 20 ns duration single cycle sinusoidal wave that changed sign every pulse. AWG also output a CW 50,004,990 Hz sinusoidal signal for driving the AOM. The beating between the ultrasound transducer and the AOM was 10 Hz and the CMOS camera ran at 40 Hz, controlled by DG2. For every 25 ms, the CMOS spent the first 10 ms for exposure and during the remaining 15 ms transferred the data to a computer. The driving signals for the ultrasound transducer and the AOM were amplified to 160 Vp-p and 30 Vp-p, respectively.

**Tissue phantom preparation and scattering coefficient measurements**

1.5 micron diameter polystyrene bead suspension (2.61 % solid) was dispersed in agar (1.5 % in H$_2$O) at 33:467 volume ratio. The sample was mixed on a shaker until it appeared uniform. Specially made metal spacers were sandwiched between two pieces of cover glass to control the phantom's thickness. The polystyrene beads in agar mixture was then injected between the cover glass to create tissue phantoms. To measure the scattering coefficient, we used a collimated laser beam to illuminate the sample. We measured the ballistic component of the transmitted light at ~ two meters distance from the tissue phantom. For this measurement, we used phantoms of thickness 0.4, 0.6, 0.8 and 1 mm. For each thickness, we prepared three phantoms. The thickness of each coverglass and the total thickness were measured with a digital micrometer to yield the precise individual phantom thickness. For each phantom, we repeated the measurement at three different positions of the phantom. The experimental data was fitted with an exponential curve to determine the scattering coefficient ($\mu_s$=6.4±0.2 /mm).

## Author Contributions

## Competing financial interests

The authors declare no competing financial interests.

## Acknowledgements

The authors thank Charles Shank, Ying Min Wang, and Changhuei Yang for helpful discussions and thank Tsai-Wen Chen for instructions on the micropipette puller. The research is supported by Howard Hughes Medical Institute.

Figure 1

(*a*) Experimental scheme of fluorescence microscopy by single cycle ultrasound pulse guided DOPC. (*b*) Experiment setup. λ/2, half wave plate; PBS, polarizing beam splitter; BB, beam block; BS, non-polarizing beam splitter; BE, beam expander; M, mirror; BP, band-pass filter; LP, long-pass filter; L1, f = 35 mm lens; L2, f = 50 mm lens; D, fluorescence detector; Stage, 3-axis motorized translation stage.

Figure 2

(*a*) Measured transverse PSF. (*b*) Measured axial PSF. (*c*) and (*d*) are the corresponding images resampled with bicubic interpolation. (*e*)-(*g*) Gaussian fitting of the measured PSF.

Figure 3

(*a*) Direct optical imaging of the fluorescence hole array without tissue phantoms. (*b*) Direct optical imaging of the fluorescence hole array surrounded by 2 mm thick tissue phantoms. (*c*) Image acquired with ultrasound pulse guided DOPC through tissue phantoms. The laser power on the sample was 25 mW during sound modulation and 10 mW during fluorescence excitation. (*d*) Bicubic interpolation of (*c*).

Figure 1

Figure 2

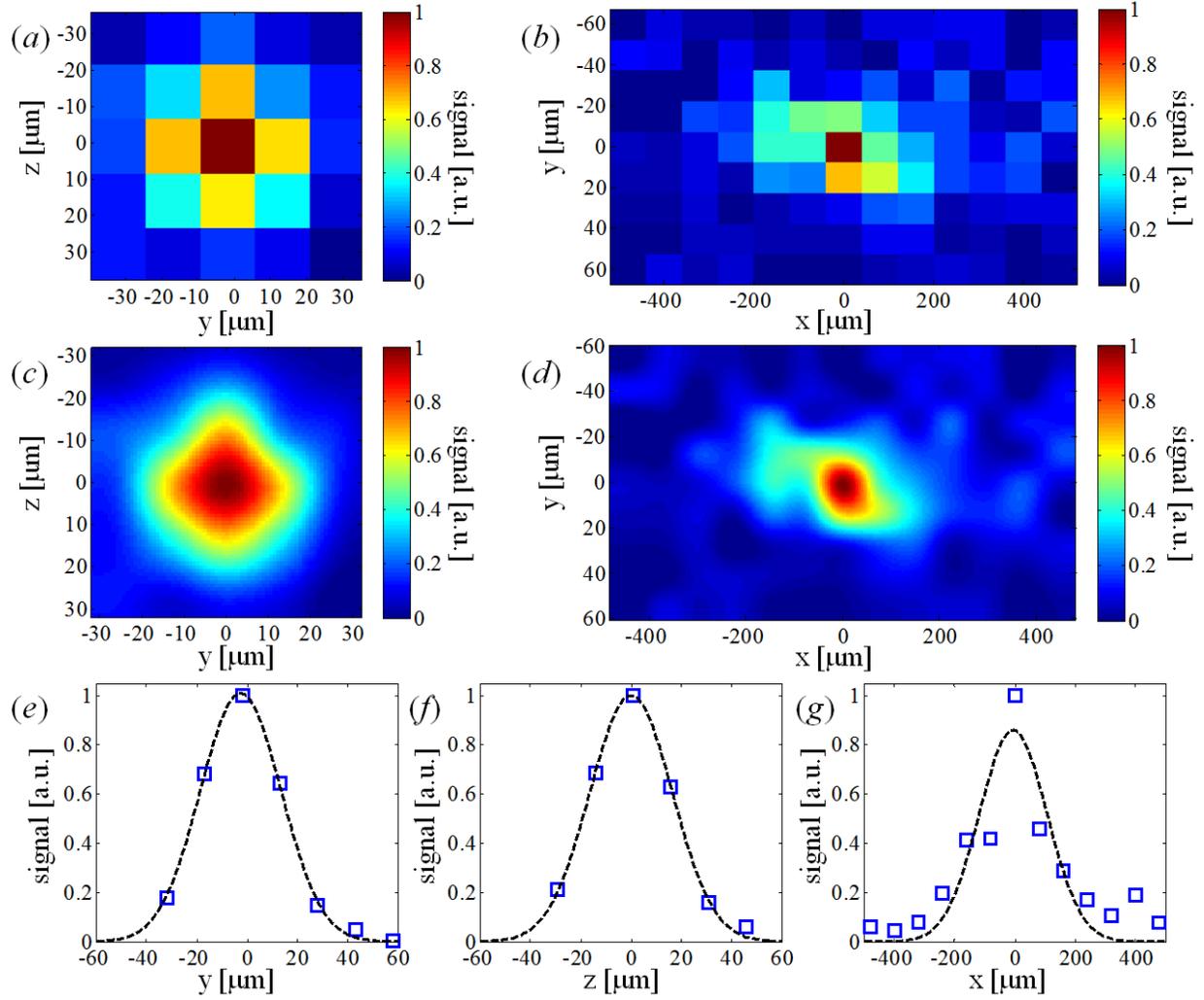

Figure 3

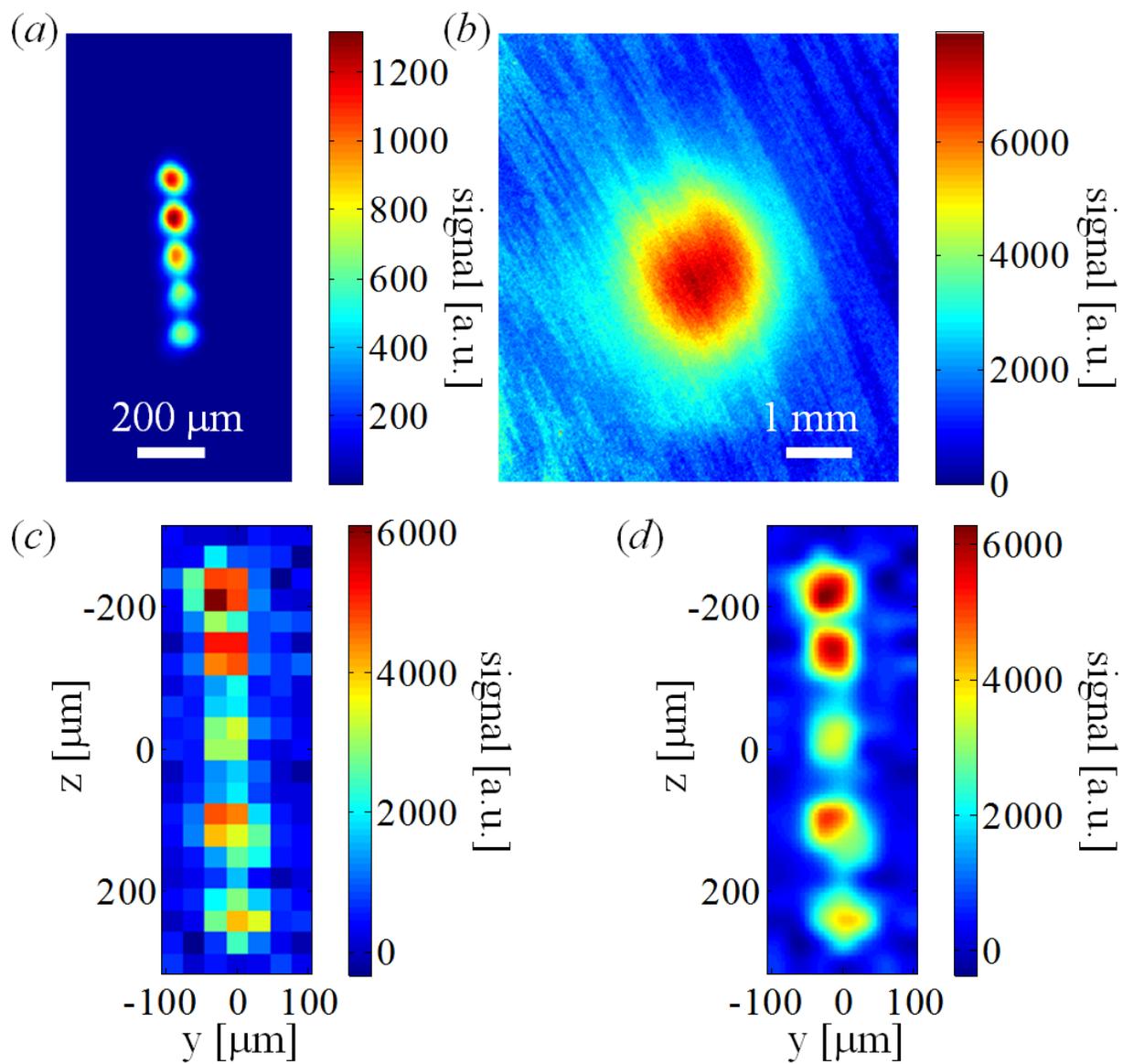

**Supplementary discussion**

Here we discuss the parameters that determine the dimensions of the sound modulation zone. The transverse confinement is determined by the focusing parameters of the sound transducer. The transducer's (Olympus NDT v3330) element diameter and focal length are 0.25 inch and 0.2 inch, respectively. The ideal full width half maxim (FWHM) of the sound focus should be $1.02F\lambda/D$, where F is the focal length, $\lambda$ is the sound wavelength and D is the element diameter. For a 50 MHz sound wave in water (sound velocity 1500 m/sec), the ideal FWHM is 24.5 microns. Supplementary Figure 3 (*a*) shows the measured dependence of the sound modulated light power on the driving voltage of the sound transducer. With low (< 50 V) driving voltage, the modulated light power shows a quadratic dependence on driving voltage or a linear dependence on the driving power ($\propto V^2$). In experiments, we used high (160 $V_{p-p}$) driving signal to achieve good modulation efficiency. With high driving voltage, the signal approaches linear dependence on the driving voltage, making the sound modulation zone larger than the sound focus. In addition, the driving signal is a 20 ns single cycle sinusoidal pulse, as shown in Supplementary Fig. 3 (*c*). Its Fourier transform (Supplementary Fig. 3 (*e*)) contains a broad bandwidth including lower frequency longer wavelength sound wave, which makes the focus larger than the value calculated with 50 MHz sound wave. The axial confinement is determined by the speed of sound, the sound pulse duration and the laser pulse duration. For laser pulses much shorter than 20 ns, the sound pulse position hardly changes during laser illumination and the axial confinement approximately equals the sound pulse duration multiplied by the speed of sound. In experiments, the laser pulse FWHM duration is 20 ns (Supplementary Fig. 3 (*b*)). During laser illumination, the sound pulse travels and the axial confinement is therefore the speed of sound multiplied by the sound pulse duration convolved with the laser illumination time.

Supplementary Figure 3 (*d*) is the convolution of Supplementary Fig. 3 (*b*) with the amplitude of Supplementary Fig. 3 (*c*). The FWHM of Supplementary Fig. 3 (*d*) is 25.5 ns and the axial confinement is therefore 38.4 microns, which is close to the measured 37.9 microns.

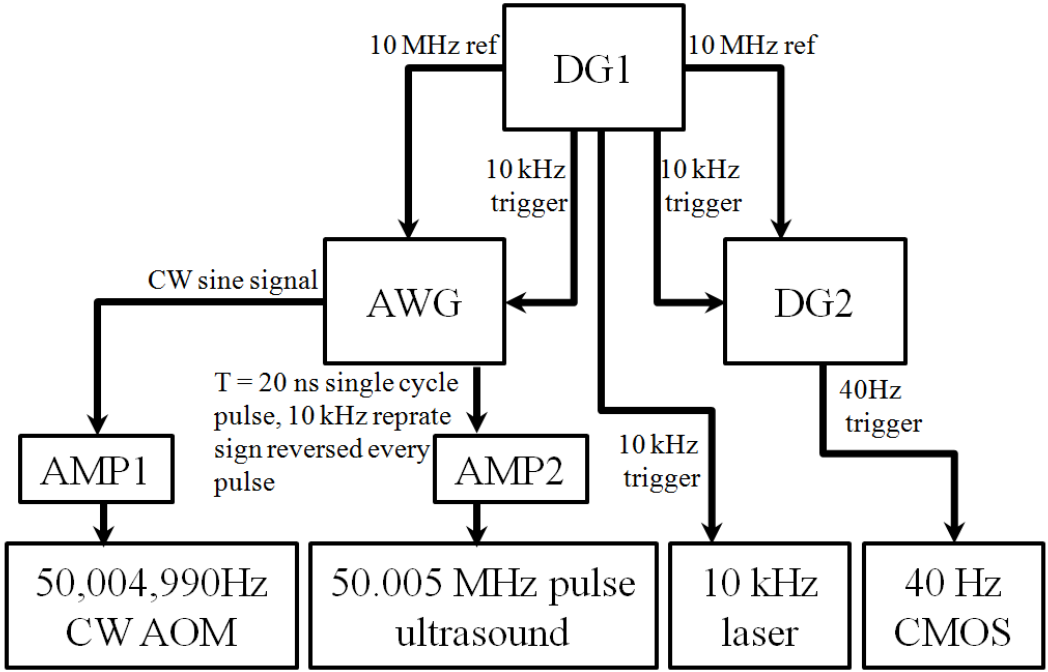

Supplementary Figure 1,

Timing and synchronization diagram. DG1, delay generator (Stanford Research DG645); AWG, arbitrary waveform generator (Tektronix AFG3252); DG2, delay generator (Stanford Research DG535); AMP1, RF amplifier (AR, 75A250A); AMP2, RF amplifier (AR, 25A250A).

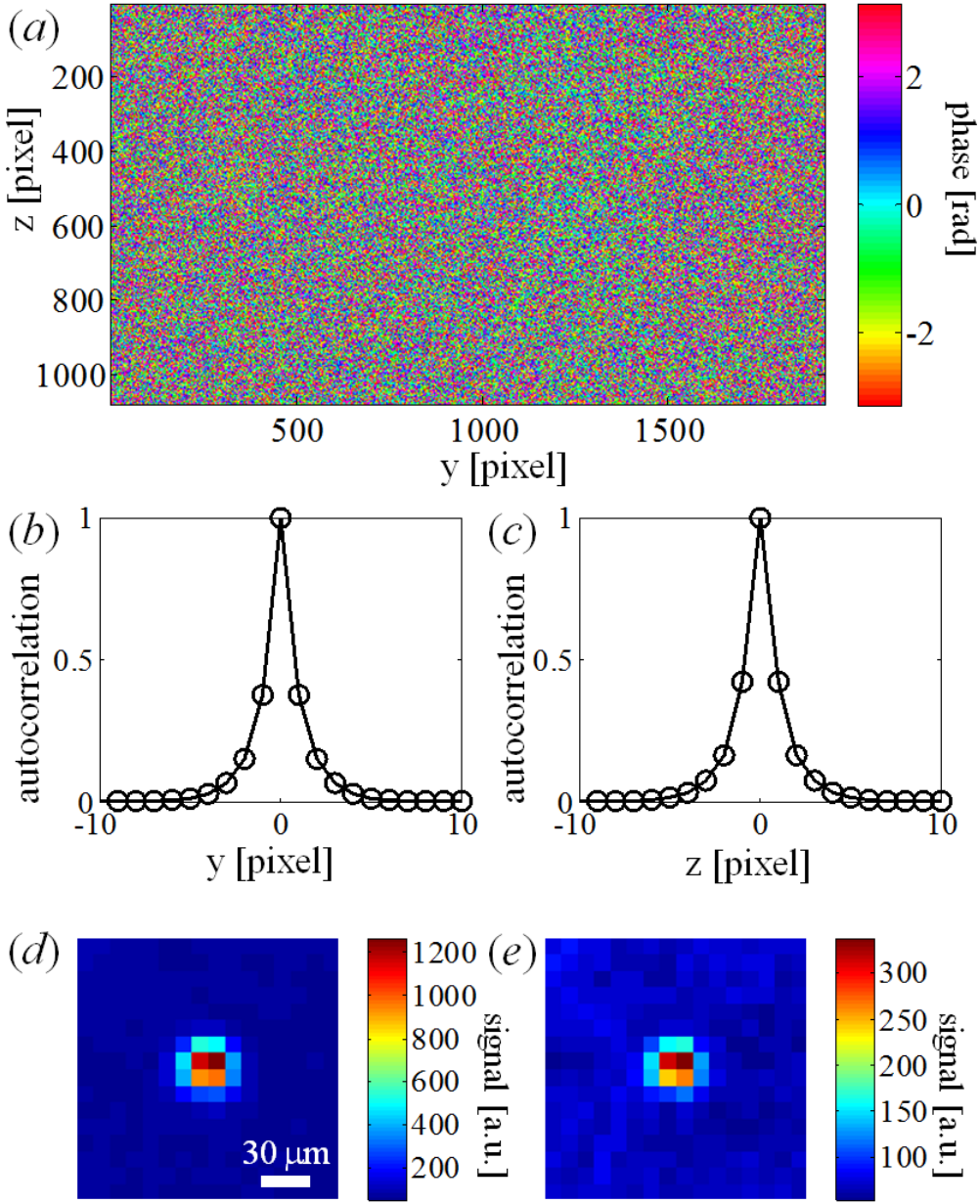

Supplementary Figure 2

(*a*) One of the phase patterns $p(y,z)$ used for phase conjugation. (*b*) and (*c*) are the autocorrelations of $\exp(ip(y,z))$ along y and z directions, respectively. (*d*) Image of one fluorescence bead illuminated by the OPC beam. (*e*) Image of the bead after the phase pattern was shifted on the SLM (making OPC ineffective).

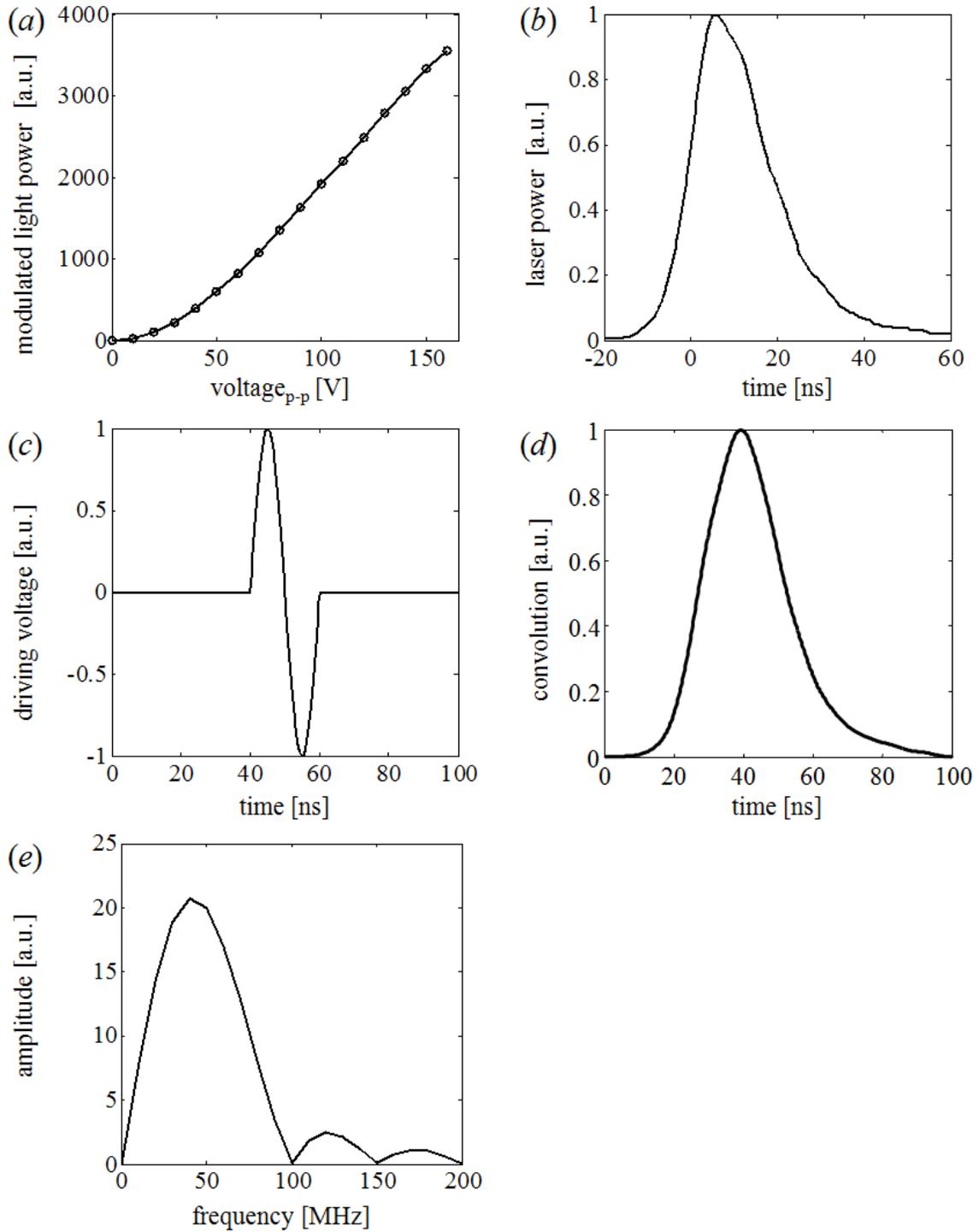

Supplementary Figure 3

(*a*) Dependence of the modulated light power on the driving voltage of the ultrasound transducer. (*b*) Measured temporal profile of the laser pulse. (*c*) Single cycle sinusoidal pulse used for driving the ultrasound transducer. (*d*) Convolution of (*b*) and (*c*). (*e*) Fourier transform of (*c*).